# A Two-step-training Deep Learning Framework for Real-time Computational Imaging without Physics Priors

RUIBO SHANG, KEVIN HOFFER-HAWLIK, AND GEOFFREY P. LUKE[*]

*Thayer School of Engineering, Dartmouth College, 14 Engineering Dr., Hanover, NH 03755, USA*
*[*]geoffrey.p.luke@dartmouth.edu*

**Abstract:** Deep learning (DL) is a powerful tool in computational imaging for many applications. A common strategy is to reconstruct a preliminary image as the input of a neural network to achieve an optimized image. Usually, the preliminary image is acquired with the prior knowledge of the imaging model. One outstanding challenge, however, is the degree to which the actual imaging model deviates from the assumed model. Model mismatches degrade the quality of the preliminary image and therefore affect the DL predictions. Furthermore, many computational imaging algorithms for the preliminary image reconstruction are time consuming and therefore cannot achieve real-time DL-based imaging. Another main challenge is that since most imaging inverse problems are ill-posed and the networks are over-parameterized, DL networks have flexibility to extract features from the data that are not directly related to the imaging model. This can lead to suboptimal training and poorer image reconstruction results. To solve these challenges, a two-step-training DL (TST-DL) framework is proposed for real-time computational imaging without physics priors. First, a single fully-connected layer (FCL) is trained to directly learn the model with the raw measurement data as input and the image as output. Then, this pre-trained FCL is fixed and concatenated with an un-trained deep convolutional network with a U-Net architecture for a second-step training to improve the output image fidelity. This approach is shown to have four main advantages. First, it does not rely on an accurate representation of the imaging model since the first-step training is to directly learn the model. Even when there is no model mismatch, the results are still comparable to those in the DL approaches which incorporate the model. Second, real-time imaging can be achieved since the raw measurement data is directly used as the input to the network. Third, the TST-DL network is trained in the desired direction and the predictions are improved since the first step is constrained to learn the model and the second step improves the result by learning the optimal regularizer. Fourth, the approach can accommodate any size and dimensionality of network input and solve the input-output size-and dimensionality- mismatch issues which arise in convolutional neural networks. We demonstrate this framework using a linear single-pixel camera imaging model. The results are quantitatively compared with those from other DL frameworks and model-based iterative optimization approaches. Robustness to model mismatch, the advantage of the two-step training, noise robustness and the required size of the training dataset are studied for this framework. We further extend this concept to nonlinear models in the application of image de-autocorrelation by using multiple FCLs in the first-step training. Overall, this TST-DL framework is a flexible approach for real-time image reconstruction without physics priors, applicable to diverse computational imaging systems.

## 1. Introduction

Computational Imaging is a powerful tool in the application of image reconstruction. It relaxes the hardware requirements of imaging systems by relying on (typically iterative) computational techniques to recover the lost information, that is, solving an inverse imaging problem computationally [1, 2]. These methods rely on a measured or assumed forward operator of the imaging system to create a mapping from the image to the measurement. However, the forward operator is often ill-posed by design or due to the imperfect physical measurement, meaning multiple solutions exist for a given measurement. Therefore, additional information about the scene or the object must be incorporated in the computational process for accurate reconstruction.

One of the most common methods in computational imaging is sparsity-based optimization which seeks to reconstruct images from incomplete data or an ill-posed forward operator [3, 4]. This concept is based on the knowledge that most natural images are sparse (i.e., only a few nonzero values exist) when transformed into a specific domain. Researchers have successfully applied sparsity-based optimization in a variety of imaging fields ranging from compressed ultrafast photography [5] to holographic video [6] to biomedical imaging [7]. Although sparsity-based optimization has advantages in image reconstruction, the primary drawback to this approach is that it is iterative and time consuming. Depending on the scale and scope of the problem, an image reconstruction task can require minutes to even hours of computation. Therefore, it cannot achieve real-time imaging for many applications which require pipelined data acquisition and image reconstruction. Furthermore, the optimal algorithm-specific parameters in the sparsity-based optimization framework must be heuristically determined.

Deep learning (DL) [2, 8] is an emerging computational imaging approach dramatically improving the state-of-the-art in fast image reconstruction [9-14]. Instead of building a specific model and finding the optimal algorithm-specific parameters heuristically (as in sparsity-based optimization approaches), it relies on large amounts of data to automatically learn tasks by using the backpropagation algorithm to find the optimal parameters in each layer of a neural network [8]. It has the benefit of being computationally efficient since most of the computational energy is used during the one-time training process. Compared with sparsity-based optimization approaches which require iterative testing of the regularizer for each image [15], DL approaches utilize the training dataset to find the optimal regularizer for a broad range of images. Therefore, DL is a promising alternative to augment or replace the iterative algorithms used in sparsity-based optimization. Researchers have applied the DL approach in many imaging fields with varying network structures [2]. The U-Net [16] architecture is one of the most successful DL frameworks in the imaging field. Its architecture consists of a contracting path to capture context and a symmetric expanding path for enhancement of key features. Skip connections between the contracting and expanding path help to preserve features from the input image. A variety of applications in the imaging field, ranging from segmentation to image reconstruction from incomplete data, have harnessed the original or a modification of the U-Net structure [17-23]. Despite the advantages of these DL approaches in computational imaging, there are still some limitations. First, most of the current DL techniques still require knowledge of the imaging model for an initial image guess to feed into the DL networks [14, 21, 24, 25]. However, the forward model in many imaging fields can be difficult to acquire with high accuracy (a model mismatch) [7, 26-28]. The model mismatch will lead to the inaccurate initial image guess and therefore affects the DL prediction. Furthermore, the reconstruction of the initial image guess will sometimes be computationally intensive, especially when using iterative image reconstruction approaches [29, 30]. Second, most of the computational imaging problems we are solving are ill-posed and the networks are over-parameterized. This means most DL networks have flexibility to extract features from the data that are not directly related to the imaging model (there are no constraints to enforce learning of the physical model rather than extraction of image features) [31]. This will make the network to be trained in the undesired direction and affect the network predictions Third, most of the DL approaches are designed for a specific application and not widely applicable in other problems. For instance, mismatches of the size and dimensionality between the measurement data and the reconstructed image are not easily addressed in the U-Net architecture which usually requires the input and output images to have the same size and dimensionality [16]. Although a modified U-Net framework can deal with the image size mismatch issues, it still requires a two-dimensional (2D) image as the input [22].

In this paper, a two-step-training DL (TST-DL) framework is proposed for real-time DL-based computational image reconstruction without prior knowledge of the forward model. The first step trains a single fully-connected layer (FCL) (for linear imaging models) or multiple FCLs (for nonlinear imaging models) to approximate the imaging inverse model with the raw measurement data as input and the ground-truth image as output. The weights of this trained FCL are then fixed and concatenated with an untrained convolutional neural network (U-Net) for a second-step training to effectively impose regularization constraints and improve the reconstruction quality of the results predicted from the first-step training. This versatile DL approach is beneficial to diverse computational imaging systems for the first reason that it does not need to consider the level of the model mismatch by directly learning the model instead of using the prior knowledge of the model (*Section 3.3* for details). Second, we decoupled the whole inverse problem into two sub-problems with the two-step-training strategy. We added constraints to enforce the first step (FCL) to learn the model and the second step to learn the optimal regularizer instead of simply extracting features in some one-step end-to-end DL approaches (*Section 3.4* for details). It also overcomes the overburdening issues stemming from learning the model and the optimal regularizer simultaneously in established end-to-end DL approaches. Furthermore, it can yield results in real time with comparable performance to the iterative algorithms. Moreover, it can handle any size or dimensionality of the network input and solve the input-output image size- and dimensionality- mismatch issues. Finally, by incorporating minor changes to the TST-DL network, image reconstruction with nonlinear imaging models can be performed.

## 2. Methods

### 2.1 Regularized optimization

Any imaging model can be described by

$$g = \boldsymbol{H}f \tag{1}$$

where $f$ is the image to be reconstructed, $g$ is the raw measurement data and $\boldsymbol{H}$ is the forward operator.

The most straightforward way to reconstruct the image $f$ is to find the inverse of the forward operator $\boldsymbol{H}^{-1}$ so that $\boldsymbol{H}^{-1}\boldsymbol{H} = \boldsymbol{I}$ where $\boldsymbol{I}$ is the identity matrix. However, for most of the cases, $\boldsymbol{H}^{-1}$ is not unique or requires excessive computational power to determine.

An effective alternative to directly computing the inverse of the forward model is to iteratively solve the optimization problem,

$$\hat{f} = \underset{f}{\operatorname{argmin}} \ \|\boldsymbol{H}f - g\|_2^2 \tag{2}$$

where $\|\cdot\|_2$ denotes the L$_2$ norm. However, this pseudo-inverse solution is prone to artifacts and noise due to the ill-posed property of the forward operator $\boldsymbol{H}$. Therefore, additional information is needed to converge to the correct solution.

A regularized optimization approach can incorporate additional knowledge about the image by adding a regularization term,

$$\hat{f} = \underset{f}{\operatorname{argmin}} \ \{\|\boldsymbol{H}f - g\|_2^2 + \lambda \phi(f)\} \tag{3}$$

where $\phi$ is the regularization operator and $\lambda$ is the regularization parameter. $\|\boldsymbol{H}f - g\|_2^2$ is the fidelity term and $\phi(f)$ is the regularization term. The regularization term is to make a balance with the fidelity term by driving the optimized $\hat{f}$ to match a specific regularization rule. The common regularization domains include spatial, edge, or wavelet domains and so on. However, finding the optimal regularization rule for a specific image dataset is still a challenging problem [15].

Inspired by the regularized optimization approach, we propose TST-DL framework. The first-step training is to train an FCL to learn an optimal $H^{-1}$ (assuming $H$ is a linear operator) given the training datasets $(g, f)$. Then, this pre-trained FCL is fixed and concatenated with a U-Net for the second-step training to learn an optimal regularization rule to regularize $f$ towards the optimal solution. By decoupling the whole inverse problem into two sub-problems with the two-step-training strategy, we enforce the first step (FCL) to learn the model and the second step to learn the optimal regularizer (*Section 3.4* for details). We further extend this concept to nonlinear models by using multiple FCLs to learn the optimal nonlinear operator $H^{-1}$.

*2.2 The TST-DL structure*

Our TST-DL framework contains an FCL (or multiple FCLs for nonlinear models) and a U-Net architecture as shown in Fig. 1. The DL framework in step 1 consists of an FCL mapping from the raw measurement data (input) to the image (output) (Batch-normalization (BatchNorm), reshape and permute layers are used for normalization and reshape purposes). With this FCL, the input measurement data and the output image do not need to have the same size or even the same dimensionality. By training the FCL, the optimal inverse operator will be learned given the training datasets. The DL framework in step 2 follows the U-Net architecture concatenated with the pre-trained FCL from step 1. The U-Net, which utilizes an encoder-decoder structure with skip connections to preserve wide-frequency features, was chosen because of its success in solving image-to-image problems. The mean squared error (MSE) is used as the loss function in the first-step training to find the optimal $H^{-1}$ that minimizes $\|f - H^{-1}g\|_2^2$. A customized loss function with a combination of root mean squared error (RMSE) and difference of structural similarity index (DSSIM) is used for the second-step training. The default learning rate is used. The batch size is chosen to be 50 and each training step runs 100 epochs. Dropout layers (not shown in Fig. 1) are used in the second step (U-Net) of TST-DL to prevent overfitting issues.

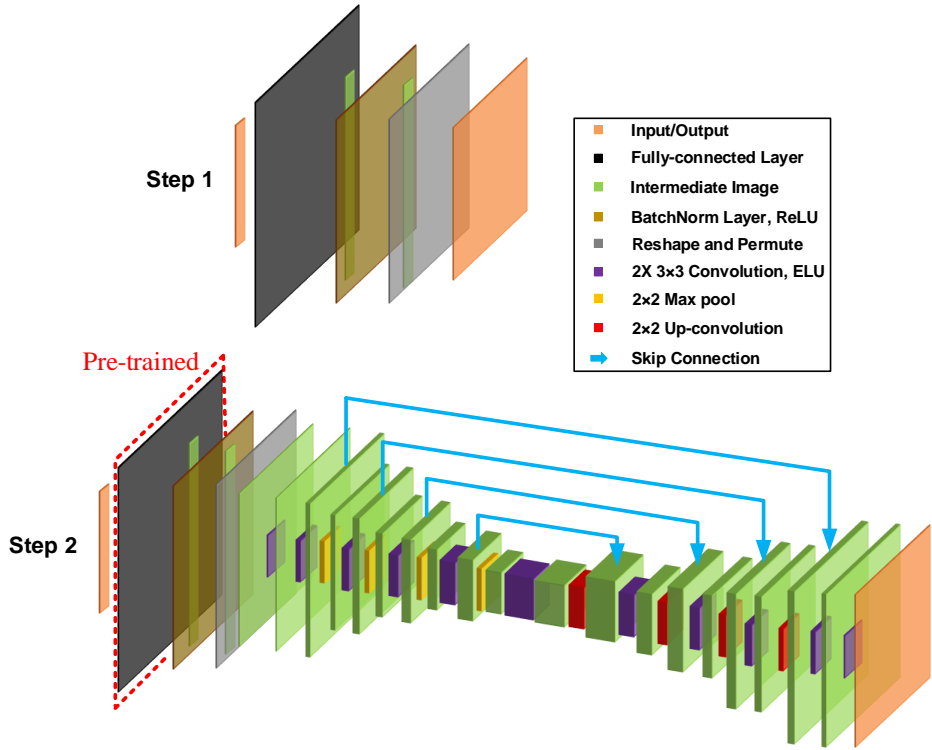

**Fig. 1**. TST-DL structure. Step 1 is training the FCL and step 2 is training a U-Net concatenated with the fixed pre-trained FCL. The input is the raw measurement data that can be any size and dimension and the output is a 2D image.

## 3. Data acquisition and results for different imaging models

The data acquisition and results from the prediction of the TST-DL for the single-pixel imaging with Russian-Doll (RD) Hadamard [32] and random Hadamard patterns, and image de-autocorrelation are shown in this section with the simulated data. The mismatch between the expected and actual model is not considered in *Section 3.1* and *3.2*. In this case, we can see if the results from TST-DL are comparable to those from the approaches that incorporate the model when no model mismatch exists. Then, the model mismatch is analyzed in *Section 3.3* in details to verify that TST-DL starts to outperform the approaches that incorporate the model as the model mismatch increases. Quantitative comparisons are made with other DL frameworks (a deep convolutional auto-encoder network (DCAN) [33], one-step-training DL (OST-DL) and U-Net) and the established model-based optimization approaches (an iterative L2 norm minimization approach LSQR [29] and a two-step iterative shrinkage/thresholding (TwIST) algorithm [30]). Note that the U-Net approach assumes accurate initial guess of the image with the knowledge of the model while the LSQR and TwIST methods require precise knowledge of the forward model. The DCAN is developed in single-pixel imaging to reconstruct the dynamic scenes from the single-pixel camera capture of the compressed signal. DCAN is comprised of two parts, the encoding part to find the optimal binary filters for the measurement and the decoding part for image reconstruction with FCL and three convolutional layers [33]. We only use the decoding part in DCAN since the binary filters as the physics priors are unknown. For the U-Net approach, an initial guess of the image is reconstructed using the LSQR approach. Then, the initial guess of the image is used as the input of U-Net for further training and prediction. For OST-DL, as an end-to-end DL approach, the FCL is concatenated with U-Net for a one-step training to learn the model and the optimal regularizer simultaneously instead of training each individually. For DCAN and OST-DL, both as one-step training approaches, the training runs 200 epochs for a fair comparison with TST-DL. For U-Net, since the initial guess of the image is obtained because of the known forward model, the training runs 100 epochs for a fair comparison with TST-DL.

### 3.1 Single-pixel imaging with RD Hadamard patterns

In the first case, the RD Hadamard patterns are used in the single-pixel imaging. In RD Hadamard patterns, the measurement order of the Hadamard basis is reordered and optimized according to their significance for general scenes, such that at discretized increments, the complete sampling for different spatial frequencies is obtained [32]. The STL-10 natural image database [34] was used for training the TST-DL framework with 10,000 images as the training dataset, 2,000 images as the validating dataset and another 2,000 images as the testing dataset. In order to meet the dimension requirement of the RD Hadamard patterns, all the images were down-sampled from 96×96 to 64×64. The full RD Hadamard basis for a 64×64 image has 4,096 RD Hadamard patterns each with a size of 64×64. Different compression ratios were used here as 4X, 8X, 16X, 32X, 64X and 128X, corresponding to taking the first 1/4, 1/8, 1/16, 1/32, 1/64, and 1/128 of RD Hadamard patterns, respectively. For instance, in 4X compression, the first 1,024 RD Hadamard patterns were used. The 1D raw measurement data was acquired by multiplying each individual image with the RD Hadamard patterns at each compression ratio. Therefore, the 1D raw measurement data has a size of 1,024×1, 512×1, 256×1, 128×1, 64×1 and 32×1 for the corresponding compression ratios.

The reconstructed images at 4X and 8X compression ratios from multiple reconstruction approaches are shown in Fig. 2. The images from the TST-DL are shown in the last column and compared with the other three DL frameworks (DCAN, OST-DL and U-Net) and the established model-based optimization approaches (LSQR and TwIST). The intermediate results from the first-step training using the FCL in TST-DL are also shown as FCL-DL.

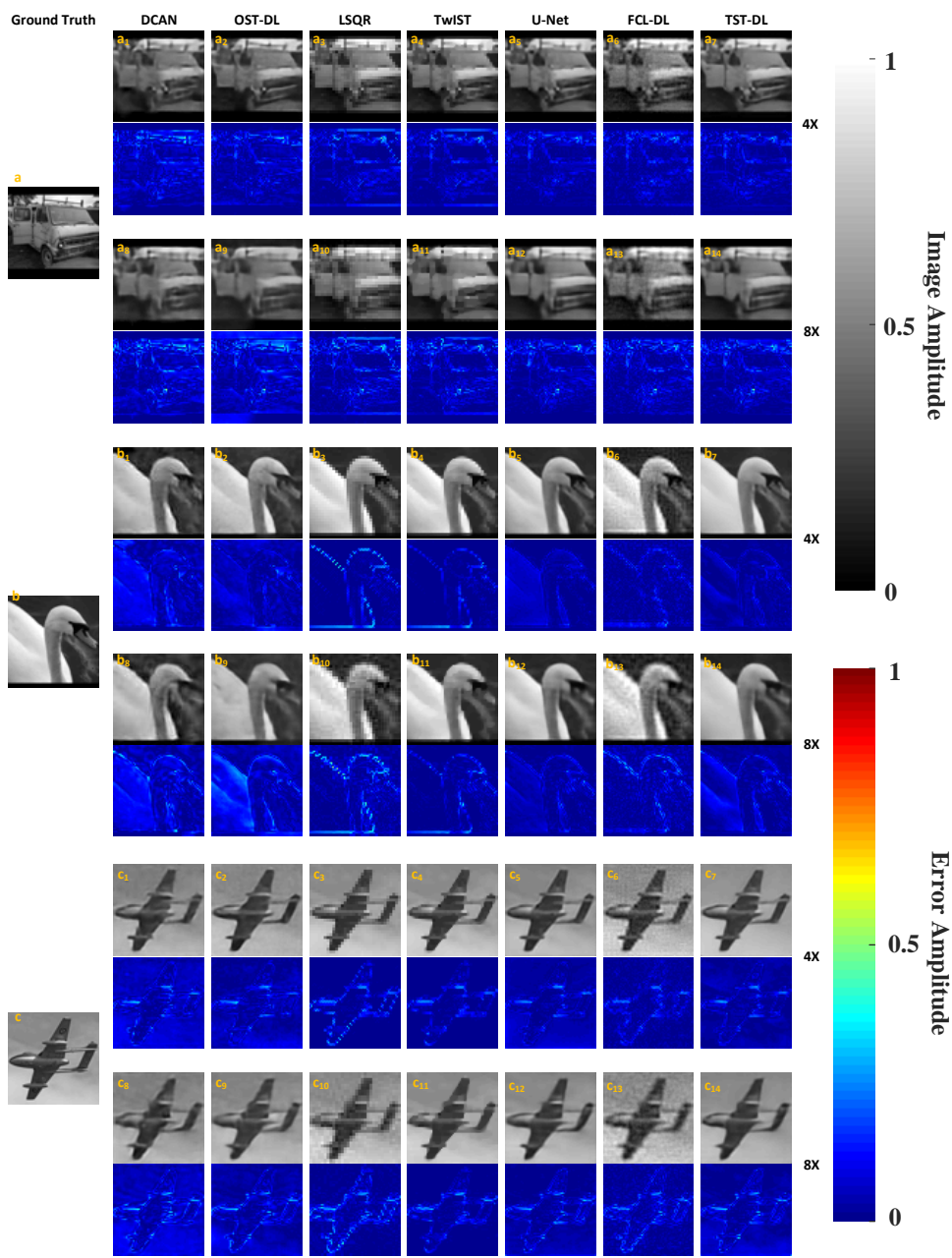

Fig. 2. Three representative reconstructed images in the testing dataset from DCAN, OST-DL, LSQR, TwIST, U-Net, FCL-DL and TST-DL at 4X and 8X compression ratios, their ground-truth images and their corresponding error images. (a) The ground-truth image of a van. ($a_1$-$a_7$) The reconstructed van images from all the approaches respectively at the 4X compression ratio and their corresponding error images. ($a_8$-$a_{14}$) The reconstructed van images from all the approaches respectively at the 8X compression ratio and their corresponding error images. (b) The ground-truth image of a swan. ($b_1$-$b_7$) The reconstructed swan images from all the approaches respectively at the 4X compression ratio and their corresponding error images. ($b_8$-$b_{14}$) The reconstructed swan images from all the approaches respectively at the 8X compression ratio and their corresponding error images. (c) The ground-truth image of an aircraft. ($c_1$-$c_7$) The reconstructed aircraft images from all the approaches respectively at the 4X compression ratio and their corresponding error images. ($c_8$-$c_{14}$) The reconstructed aircraft images from all the approaches respectively at the 8X compression ratio and their corresponding error images.

Table 1 shows the averaged root mean squared error (RMSE) and structural similarity index (SSIM) [35] of the reconstructed images in the testing dataset at 4X and 8X compression ratios from all the reconstruction approaches (For TwIST, 500 images in the testing dataset were reconstructed and used to calculate the averaged RMSE and SSIM instead of the full testing dataset in the interest of time). From Fig. 2 and Table 1, we can see that the results from DL approaches are comparable to those from the established model-based optimization approaches (LSQR and TwIST). We also show the per-pixel accuracy from the error images where the difference between the ground-truth image and the corresponding reconstructed image is calculated. We can see that most of the errors come from the edges of the images. This is expected since high frequency information is often the most difficult to reconstruct in compressed sensing applications. Importantly, both LSQR and TwIST approaches require accurate knowledge of the forward model for image optimization. In addition, reconstruction from LSQR and TwIST require thousands of iterations, which cannot achieve real-time imaging. Therefore, for real-time imaging, TST-DL is the better option than LSQR and TwIST.

Table 1. Averaged RMSE and SSIM of the reconstructed images in the testing dataset at 4X and 8X compression ratios for different approaches in single-pixel imaging with RD Hadamard patterns.

| Compression Ratio | | DCAN | OST-DL | LSQR | TwIST | U-Net | FCL-DL | TST-DL |
|---|---|---|---|---|---|---|---|---|
| 4X | RMSE | 0.079 | 0.059 | 0.060 | 0.051 | 0.050 | 0.051 | 0.051 |
|  | SSIM | 0.798 | 0.803 | 0.830 | 0.865 | 0.865 | 0.816 | 0.833 |
| 8X | RMSE | 0.086 | 0.079 | 0.072 | 0.061 | 0.059 | 0.063 | 0.063 |
|  | SSIM | 0.711 | 0.700 | 0.706 | 0.780 | 0.787 | 0.727 | 0.762 |

For further quantitative comparison with the other three DL frameworks, the mean and the standard deviation of the RMSE and SSIM are calculated through the 2,000 testing images at all the compression ratios as shown in Fig. 3. We can see that for most of the cases, the U-Net approach is the best. This makes sense since the initial guess of the input images in U-Net needs the physics priors of the model. It is reasonable that the reconstruction results will be better when the exact model (with no model mismatch) is incorporated in the framework. For TST-DL, even though the physics priors of the model are unknown, the results are almost equivalent to those from U-Net and outperform those from DCAN and OST-DL. Therefore, TST-DL is the best among the approaches that do not incorporate the model and comparable to the approaches with the prior knowledge of the model.

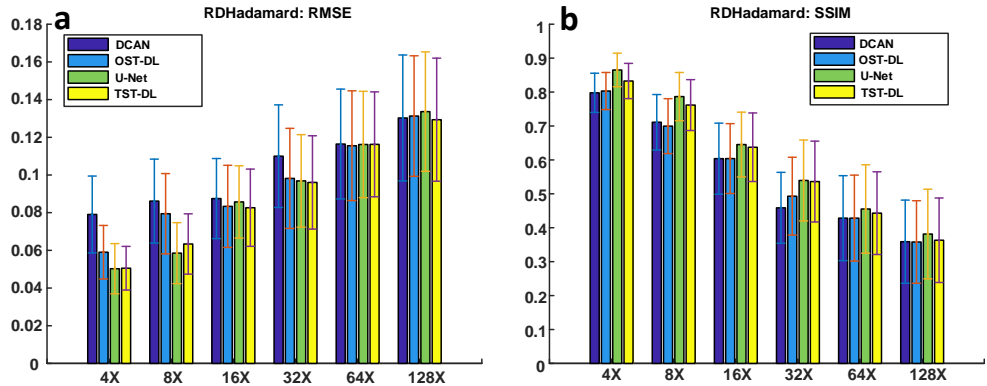

Fig. 3. Quantitative comparison in terms of RMSE and SSIM. (a) RMSE for DCAN, OST-DL, U-Net and TST-DL in all compression ratios with RD Hadamard patterns. (b) SSIM for DCAN, OST-DL, U-Net and TST-DL in all compression ratios with RD Hadamard patterns.

## 3.2 Single-pixel imaging with random Hadamard patterns

In the second case, random Hadamard patterns are used in the single-pixel imaging to test how TST-DL performs in a more challenging case since the reconstruction is more difficult than that from the RD Hadamard patterns [36]. The random Hadamard patterns were generated in Matlab by randomly permuting a full basis of 4,096 64×64-pixel patterns. The same training, validating and testing datasets were used here as in the RD Hadamard case. For random Hadamard patterns, we only reconstruct the images at 4X and 8X compression ratios.

Figure 4 shows representative results at the 4X compression ratio from DCAN, OST-DL, U-Net, TST-DL, and the ground-truth images. Visually, the TST-DL performs better than the other 3 DL frameworks. The fine details of the images can be reconstructed from TST-DL while they are only partially reconstructed or totally lost in other DL frameworks. For instance, in Fig. 4(c1-c4), the wheel of the van can be fully reconstructed in TST-DL while it is partially reconstructed in U-Net and totally lost in DCAN, OST-DL. We also show include the error images of the zoom in features. We can see that the errors are more pronounced than those using the Russian-doll Hadamard matrix, which is expected. However, in general, TST-DL still performs better than the alternative reconstruction strategies, especially for the aircraft and truck images (Fig. 4(a) and (e)).

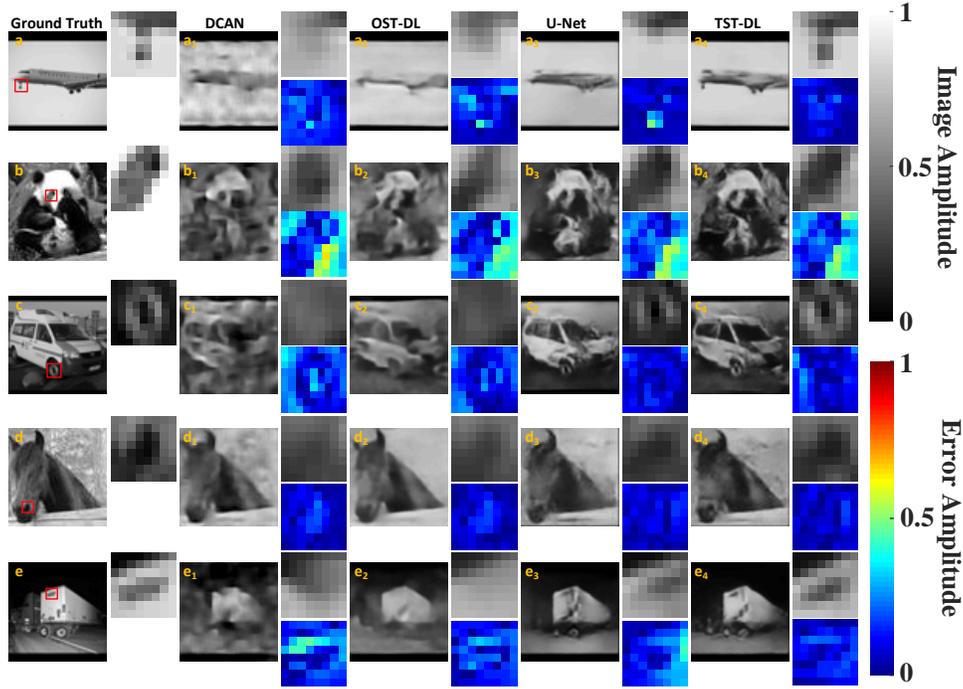

Fig. 4. Five representative reconstructed images in the testing dataset from DCAN, OST-DL, U-Net, TST-DL, some fine details with the random Hadamard patterns at the 4X compression ratio, ground-truth images and the error images of the fine details. (a) The ground-truth image of an aircraft and the fine detail of the undercarriage. ($a_1$-$a_4$) The reconstructed aircraft images from DCAN, OST-DL, U-Net and TST-DL with the fine detail of the undercarriage and their error images. (b) The ground-truth image of a panda and the fine detail of the eye. ($b_1$-$b_4$) The reconstructed panda images from DCAN, OST-DL, U-Net and TST-DL with the fine detail of the eye and their error images. (c) The ground-truth image of a van and the fine detail of the wheel. ($c_1$-$c_4$) The reconstructed van images from DCAN, OST-DL, U-Net and TST-DL with the fine detail of the wheel and their error images. (d) The ground-truth image of a horse with the fine detail of the noise. ($d_1$-$d_4$) The reconstructed horse images from DCAN, OST-DL, U-Net and TST-DL with the fine detail of the nose and their error images. (e) The ground-truth image of a truck and the fine detail of the label. ($e_1$-$e_4$) The reconstructed truck images from DCAN, OST-DL, U-Net and TST-DL with the fine detail of the label and their error images.

To quantitatively compare the results, the mean and the standard deviation of the RMSE and SSIM in the testing dataset were calculated for all the 4 DL frameworks as shown in Fig. 5 at 4X and 8X compression ratios. It shows that TST-DL performs better than the other 3 DL frameworks with lower RMSE and higher SSIM. Specifically, TST-DL performs much better than OST-DL since TST-DL learns the model and the optimal regularizer individually while OST-DL learns both simultaneously (*Section 3.4* for details). And interestingly, TST-DL is even better than the U-Net approach which requires the knowledge of the model for an initial guess of the image. The reason is that the random Hadamard matrix preserves less information of natural scenes than RD Hadamard matrix [32] so that the initial guess of the image does not include key features in the image. However, in TST-DL, there is more flexibility to incorporate features from the data in process. This makes the image reconstruction results more robust to highly ill-posed inverse models. Thus, we expect that as the information-preserving ability of the forward model degrades, the TST-DL approach will further improve.

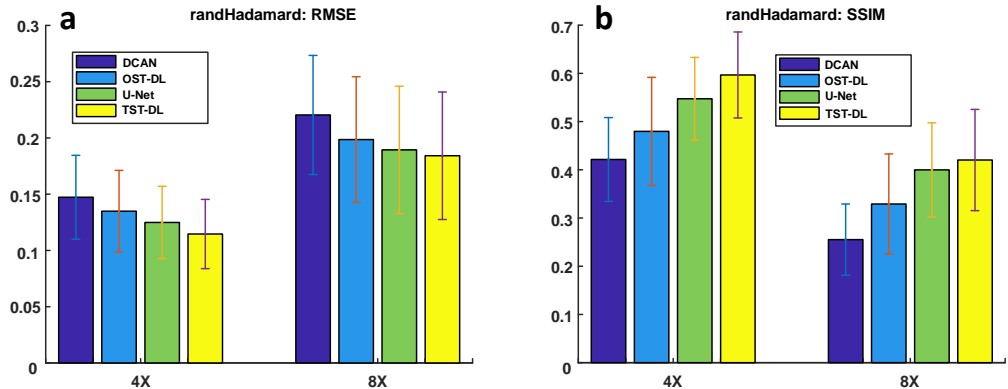

Fig. 5. Quantitative comparison in terms of RMSE and SSIM in single-pixel imaging with random Hadamard patterns. (a) RMSE for DCAN, OST-DL, U-Net and TST-DL with the random Hadamard patterns at 4X and 8X compression ratios. (b) SSIM for DCAN, OST-DL, U-Net and TST-DL with the random Hadamard patterns at 4X and 8X compression ratios.

*3.3 Robustness to model mismatch*

In the case of no model mismatch, *Section 3.1* and *3.2* show that the results from TST-DL are comparable to those from the DL and model-based iterative optimization approaches using the prior knowledge of the model. Here, we sought to explore on how the model mismatch affects the image prediction in TST-DL and the approaches that need the physics priors in single-pixel imaging with the Russian-doll Hadamard matrix. Two types of model mismatches were generated here. In the first case we randomly inverted a subset of the elements in the Russian-doll Hadamard matrix and used the modified matrix to generate the measurement data. In the second case we added different levels of uncertainty (as modeled by additive Gaussian random variables) to the Russian-doll Hadamard matrix and used the modified matrix to generate the measurement data. For image reconstruction using U-Net, we still used the original unmodified Russian-doll Hadamard matrix for an initial guess of the image as the input of U-Net. Therefore, there was a model mismatch between the data generation and the image reconstruction using U-Net. Fig. 6 shows the RMSE and SSIM results in U-Net and TST-DL with each type of the model mismatch. The results show that with the mismatch increases, TST-DL is more robust to the mismatch than U-Net. In fact, the decrease in performance stems from the fact that the actual forward model deviates from the more-ideal Russian-doll matrix. While the U-Net approach which incorporates physics priors slightly outperforms TST-DL when no model mismatch exists, its performance begins to wane as the model mismatch grows. Therefore, no matter what level the model mismatch is, one can always utilize TST-DL to learn the model through training and expect the results comparable to (no model mismatch) or better than (with model mismatch) the DL approaches that incorporate the prior knowledge of the model.

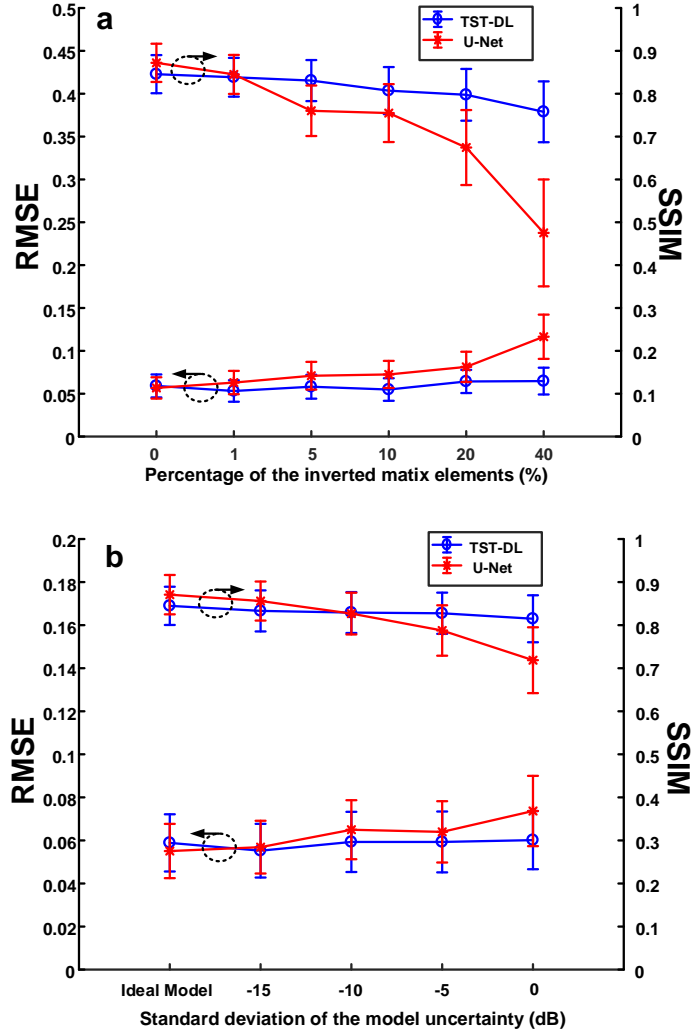

Fig. 6. Robustness to the model mismatch. (a) RMSE and SSIM of the TST-DL and U-Net results with different percentages of the inverted matrix elements. (b) RMSE and SSIM of the TST-DL and U-Net results with Gaussian random variables with varying standard deviation added to the model.

*3.4 Advantages of the two-step training strategy*

As shown in S*ection 3.1* and *3.2*, the TST-DL approach outperforms OST-DL. We hypothesized that the improved performance comes from the added constraints applied by the two-step training strategy. In this case, we are constraining the FCL to be a good approximation of the inverse model by training it alone. In order to investigate this, we analyzed the intermediate image that is produced between the FCL and the U-Net as shown in Fig. 7. We applied both the two-step (TST-DL) and the one-step (OST-DL) training strategies to the reconstruction of simulated single-pixel camera images with the random Hadamard matrix at the 4X compression ratio. In the TST-DL case, the FCL generates a relatively good approximation of the image as shown in Fig. 7(a) and (b) with the corresponding RMSE and SSIM and in Fig. 7(f) with a representative reconstructed image. Then, the U-Net effectively denoises and regularizes the estimate as shown in Fig. 7(a) and (b) with the corresponding RMSE and SSIM and in Fig. 7(g) with a representative reconstructed image. In the OST-DL case, the image after the FCL bears little resemblance to the ground truth as shown in Fig. 7(a) and (b) with a much higher RMSE and lower SSIM and in Fig. 7(d) with a representative reconstructed image. Thus, the FCL in OST-DL is learning something other than the inverse of the physical imaging model. The U-Net does a good job of completing the image reconstruction process as shown in Fig. 7(a) and (b) with the corresponding RMSE and SSIM and in Fig. 7(e) with a representative reconstructed image, but it is not quite able to achieve the performance of the TST-DL approach.

As further evidence that the OST-DL approach is failing to learn the physical model, we observed that the method was prone to overfitting to the training data as shown in Fig. 7(h) During training, the performance of prediction on the training dataset far outperformed the testing dataset as shown by the arrows for OST-DL in Fig. 7(h). This was not evident in the TST-DL case as shown by the arrows for TST-DL in Fig. 7(h). Taken together, these two findings indicate that the network architecture is complex enough to learn something other than the physical imaging model. In the absence of additional constraints, it tends to do just that.

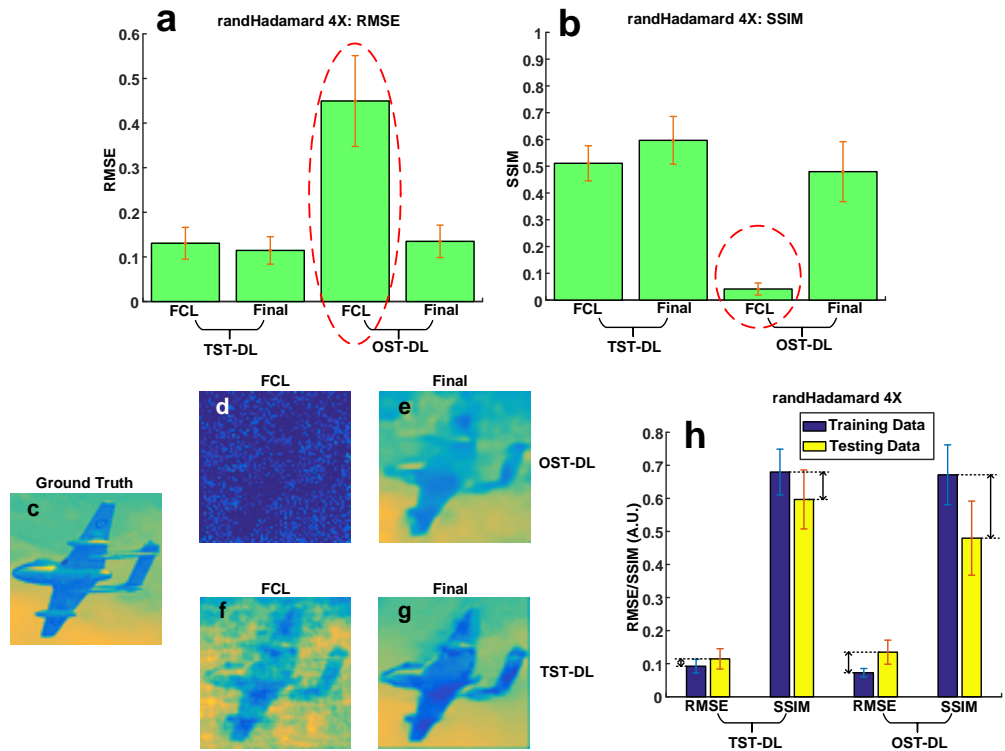

Fig. 7. Comparison between TST-DL and OST-DL with the random Hadamard matrix at the 4X compression ratio. (a) RMSE of the intermediate results after the FCL and the final results in both TST-DL and OST-DL. (b) SSIM of the intermediate results after the FCL and the final results in both TST-DL and OST-DL. (c) Ground truth of a representative image. (d) The intermediate reconstructed image after FCL in OST-DL. (e) The final reconstructed image in OST-DL. (f) The intermediate reconstructed image after FCL in TST-DL. (g) The final reconstructed image in TST-DL. (h) The RMSE and SSIM of the final results from both the training data and testing data in TST-DL and OST-DL for overfitting analysis.

## 3.5 Noise robustness

Since noise exists in the measurement data for most of the real cases, different levels of white Gaussian noise (-5dB, 0dB, 5dB, 10dB and 15dB signal-to-noise (SNR) levels) are added to the one-dimensional (1D) measurement data in single-pixel imaging with RD Hadamard patterns at the 4X compression ratio to test the robustness of the TST-DL to noise. The mean and standard deviation of the RMSE and SSIM for all the reconstructed images in the testing dataset at each noise level are calculated to quantitatively compare the performance as shown in Fig. 8 (a). It shows that with the increase of the noise level (decrease of the SNR), the reconstruction performance is dropping with the increase of the RMSE and decrease of the SSIM. However, the results still remain at a reasonable level with the RMSE lower than 0.11 and SSIM larger than 0.50 at the -5dB SNR level. Figure 8 (b-c) show the 1D measurement data without noise and with -5dB SNR of noise, respectively.

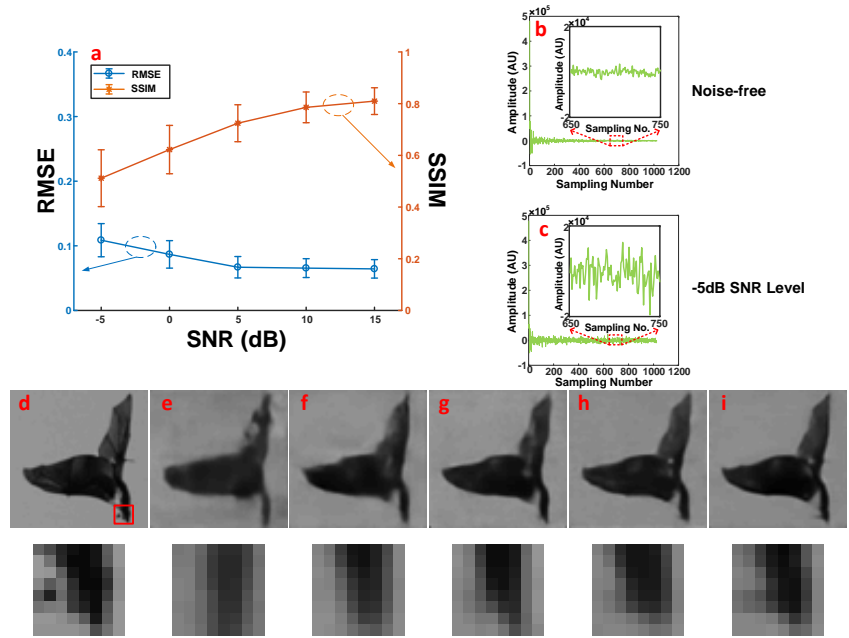

Fig. 8. Noise robustness test. (a) RMSE and SSIM of the TST-DL results at the -5dB, 0dB, 5dB, 10dB and 15dB SNR levels. (b) 1D measurement data without noise. (c) 1D measurement data at the -5dB SNR level. (d) The ground-truth of an image in the testing dataset and the fine detail in the red square. (e) The TST-DL prediction of the image and the fine detail in (d) at the -5dB SNR level. (f) The TST-DL prediction of the image and the fine detail in (d) at the 0dB SNR level. (g) The TST-DL prediction of the image and the fine detail in (d) at the 5dB SNR level. (h) The TST-DL prediction of the image and the fine detail in (d) at the 10dB SNR level. (i) The TST-DL prediction of the image and the fine detail in (d) at the 15dB SNR level.

Figure 8 (e-i) show the reconstructed images at -5dB, 0dB, 5dB, 10dB and 15dB SNR levels with the same ground-truth image in Fig. 8 (d). Although the reconstructed images become more and more blurred as the noise level increases, the general shape and even some of the details (the mouth of the bird in the zoom-in figures) can still be well reconstructed at the 0dB, 5dB, 10dB and 15dB SNR levels. Given the SNR levels at -5dB, 0dB, 5dB are extremely high levels of noise (for 0dB, the noise level is the same as the signal level), we can conclude that TST-DL is robust to noise.

## 3.6 Reducing the size of the training dataset

Because a large training dataset is not always available for real cases, the size of the training dataset is also a key factor in DL frameworks. Therefore, we test the impact of the size of the training dataset in the TST-DL to find a reasonable size of the training dataset while still maintaining good reconstruction results.

Figure 9 (a) shows the TST-DL performance of the prediction in the same testing dataset with the RD Hadamard patterns at the 4X compression ratio in terms of RMSE and SSIM with 625, 1,250, 2,500, 5,000 and 10,000 training images. The results show that with the decrease of the number of training samples, the TST-DL performance drops but still remains reasonably good at the case of 2,500 training images. Figure 9 (c-g) show the reconstruction results of the same image in the testing dataset with 625, 1,250, 2,500, 5,000 and 10,000 training images respectively together with the ground-truth image in Fig. 9 (b). The image becomes clearer and the detail is better reconstructed with the increase of the number of training samples. Qualitatively, the case of 2,500 training images has a reasonably good reconstruction result, which is consistent with the quantitative results in Fig. 9 (a). Overall, with these results, it is evident that TST-DL can still perform well with a small training dataset which means that it can be applicable to many real image reconstruction cases where a large training dataset is not easily attained.

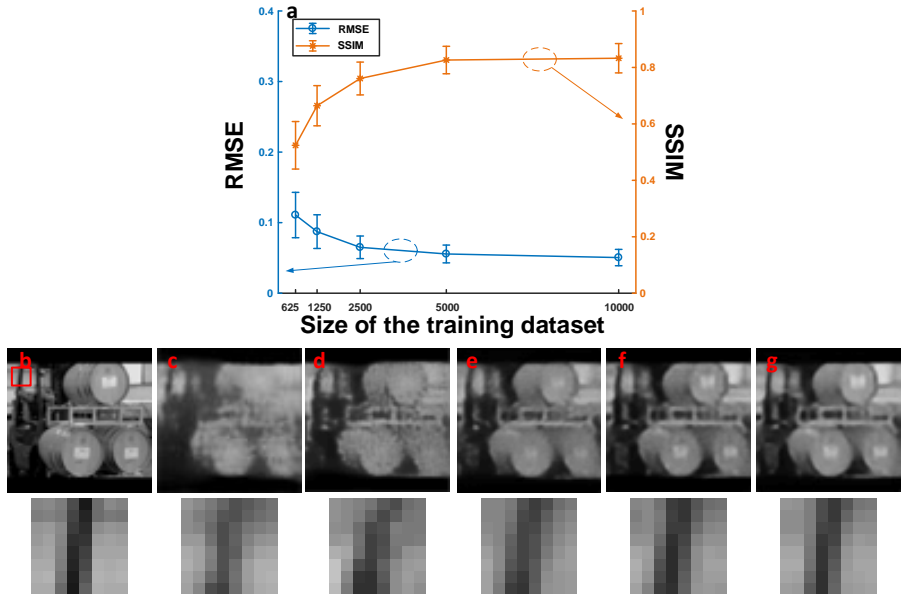

Fig. 9. Reducing the size of the training dataset. (a) The RMSE and SSIM of the TST-DL results with 625, 1,250, 2,500, 5,000 and 10,000 training images. (b) The ground truth of the oil-tank image in the testing dataset and fine detail in the red square. (c) The TST-DL prediction of the image and the fine detail in (b) with 625 training images. (d) The TST-DL prediction of the image and the fine detail in (b) with 1,250 training images. (e) The TST-DL prediction of the image and fine detail in (b) with 2,500 training images. (f) The TST-DL prediction of the image and the fine detail in (b) with 5,000 training images. (g) The TST-DL prediction of the image and the fine detail in (b) with 10,000 training images.

*3.7 Image de-autocorrelation as a nonlinear model*

For all the previous reconstruction cases, the imaging model is linear such that the forward operator can be described as a 2D matrix. Thus, the forward model and its inverse can both be implemented with matrix multiplication. TST-DL is effective at handling these imaging models since the FCL in the first-step training corresponds to matrix multiplication. The incorporation of the physics priors of a nonlinear model in a neural network is a difficult proposition. Alternatively, linearization of the model would lead to model mismatch errors described previously. Here, we extend the utility of TST-DL to handle nonlinear imaging models which cannot be described as matrix multiplication. Image autocorrelation is a nonlinear model such that the inverse process, image de-autocorrelation is also a nonlinear process. One of the important applications of image de-autocorrelation is to reconstruct the image through scattering medium [37] by solving a phase-retrieval problem from the Fourier-domain magnitude measurement [38, 39]. Therefore, we applied TST-DL to the image de-autocorrelation problem as a test case for the nonlinear model. In order to handle nonlinear models, a slight modification is made to TST-DL by using 3 FCLs connected in series instead of a single FCL in the first-step training. This additional FCLs with a nonlinear activation function add more nonlinearity to the DL network to handle the nonlinear inverse problem. The handwriting numbers in the MNIST database [40] were used as the ground-truth images. The raw images in MNIST were resized from 28×28 to 64×64 pixels. 10,000 images from the training dataset in MNIST were used as the training dataset, 2,000 images from the testing dataset in MNIST were used as the validating dataset and another 2,000 images from the testing dataset in MNIST were used as the testing dataset. Then, the image autocorrelation was applied to each of the images. The vectorized autocorrelated images were used as the input of TST-DL and their corresponding ground-truth images were used as the output of the network to achieve image de-autocorrelation. Each step in TST-DL ran 50 epochs.

Previously [37], the image de-autocorrelation was achieved through phase-retrieval algorithms. Therefore, we compare the results with those from the Gerchberg-Saxton phase-retrieval algorithm [38]. Figure 10 shows the reconstruction results where the images in Fig. 10 (a-j) are the autocorrelation images, Fig. 10 ($a_1$-$j_1$) are the ground-truth images, Fig. 10 ($a_2$-$j_2$) are the reconstructed images from the phase-retrieval algorithm, Fig. 10 ($a_3$-$j_3$) are the intermediate reconstructed images from the first-step training in TST-DL and Fig. 10 ($a_4$-$j_4$) are the reconstructed images from the TST-DL. From the results, it is evident that the phase-retrieval algorithm sometimes fails to work because of twinned image artifacts in the reconstructed images while TST-DL is much more robust. For further quantitative comparison, the RMSE and SSIM were calculated. The RMSE for all the reconstructed images in the testing dataset is 0.137 for TST-DL and 0.155 for the phase-retrieval algorithm. The corresponding SSIM is 0.815 for TST-DL and 0.735 for the phase-retrieval algorithm. Therefore, TST-DL has a more stable reconstruction than the phase-retrieval algorithm and performs better in terms of RMSE and SSIM. It also means that TST-DL is able to handle a nonlinear inverse imaging problem with a slight modification.

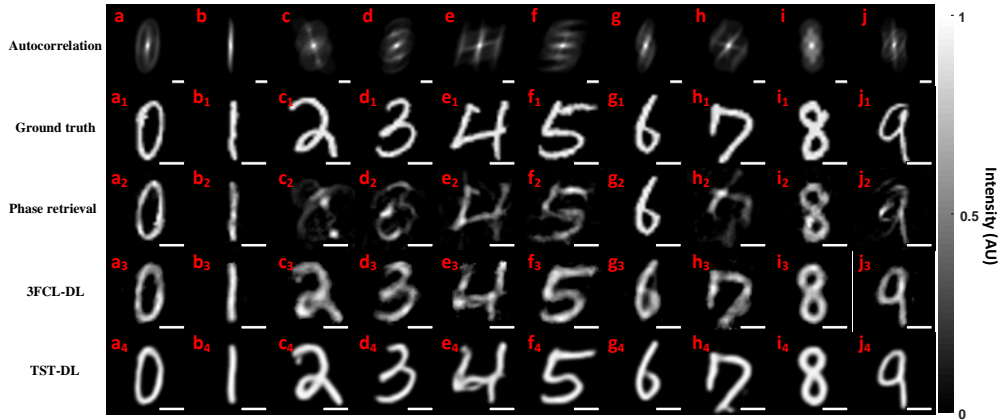

Fig. 10. Image de-autocorrelation results of ten representative images in the testing dataset from MNIST. (a-j) Autocorrelated images. ($a_1$-$j_1$) Ground-truth images. ($a_2$-$j_2$) Reconstructed images from the phase-retrieval algorithm. ($a_3$-$j_3$) Intermediate reconstructed images from the first-step training in TST-DL. ($a_4$-$j_4$) Reconstructed images from TST-DL. The scale bar denotes 20 pixels.

## 4. Experimental results in single-pixel imaging

The experimentally recorded data in single-pixel imaging with random grayscale illumination patterns was used to verify the effectiveness of TST-DL in experimental single-pixel imaging. Since TST-DL and OST-DL perform better than DCAN in the approaches without the physics priors, and U-Net performs better than TwIST and LSQR in the approaches with the physics priors, we only tested and compared TST-DL, OST-DL and U-Net with the experimental data. The objects images were taken from MNIST database [40] and resized from 28×28 to 32×32 pixels. 1,024 random grayscale illumination patterns each with a size of 32×32 were prepared as the full basis. Then, the first 51 and 256 illumination patterns in the full basis were used to illuminate the objects respectively, corresponding to a 20X and 4X compression ratio respectively. Therefore, the corresponding 1D raw measurement data has a size of 51×1 and 256×1 respectively (The 1D raw measurement data with the full illumination patterns has a size of 1,024×1). TST-DL and OST-DL were trained with 200 images in the training dataset of MNIST and their corresponding simulated 1D measurement data at the two compression ratios respectively and tested on the experimentally acquired 1D measurement data of 10 images in MNIST database different from the 200 training images. In U-Net, the reconstructed images from LSQR approach were used as the network input instead of the 1D measurement data for both training and testing cases. The batch size was chosen to be 20 since only 200 images and the corresponding 1D raw measurement data were trained. For TST-DL, each step was trained with 500 epochs. For fair comparison, OST-DL was trained with 1,000 epochs and U-Net was trained with 500 epochs since the prior knowledge of the model was provided. The imaging system is shown in Fig. 10 in [41] and in Fig. 4 in [42]. Each testing image was printed on a paper card and illuminated by the same patterns as in the simulation from a JmGO G3 projector. The 1D measurement data was recorded by a Thorlabs FDS1010 photodiode detector and a NI USB-6216 data acquisition card.

The results are shown in Fig. 11 and Table 2. Figure 11 (a-j) show 10 ground-truth images from the testing dataset. Both the intermediate images after the FCL and the final reconstructed images from OST-DL, U-Net and TST-DL at the two compression ratios are shown in Fig. 11. Quantitative comparison is made by calculating the RMSE and SSIM between the final reconstructed images and the ground-truth images as shown in Table 2. Both the figures and the quantitative comparison in Table 2 (in terms of RMSE and SSIM) show that TST-DL still outperforms OST-DL at both 4X and 20X compression ratios. This is because constraints are added to the FCL (first step) in TST-DL to enforce learning of the model while the FCL in OST-DL (with no constraint) does not learn the model. Thus, it is seen in the experiment that adding constraints to enforce learning of the model can improve the DL predictions. The U-Net and TST-DL approaches yield comparable results at a 4X compression ratio. At a 20X compression ratio, however, TST-DL is a slightly better than U-Net since the initial LSQR image reconstruction in U-Net at 20X has poor image quality and is not able to incorporate sufficient information from the training dataset to offset the high compression of the image acquisition, as seen in Fig. 11($a_9$-$j_9$).

Table 2. Averaged RMSE and SSIM of the reconstructed images at 4X and 20X compression ratios for OST-DL, U-Net and TST-DL in experimental single-pixel imaging with random grayscale illumination patterns.

| Compression Ratio | | OST-DL | U-Net | TST-DL |
|---|---|---|---|---|
| 4X | RMSE | 0.274 | 0.111 | 0.115 |
| | SSIM | 0.540 | 0.870 | 0.860 |
| 20X | RMSE | 0.293 | 0.175 | 0.161 |
| | SSIM | 0.446 | 0.733 | 0.772 |

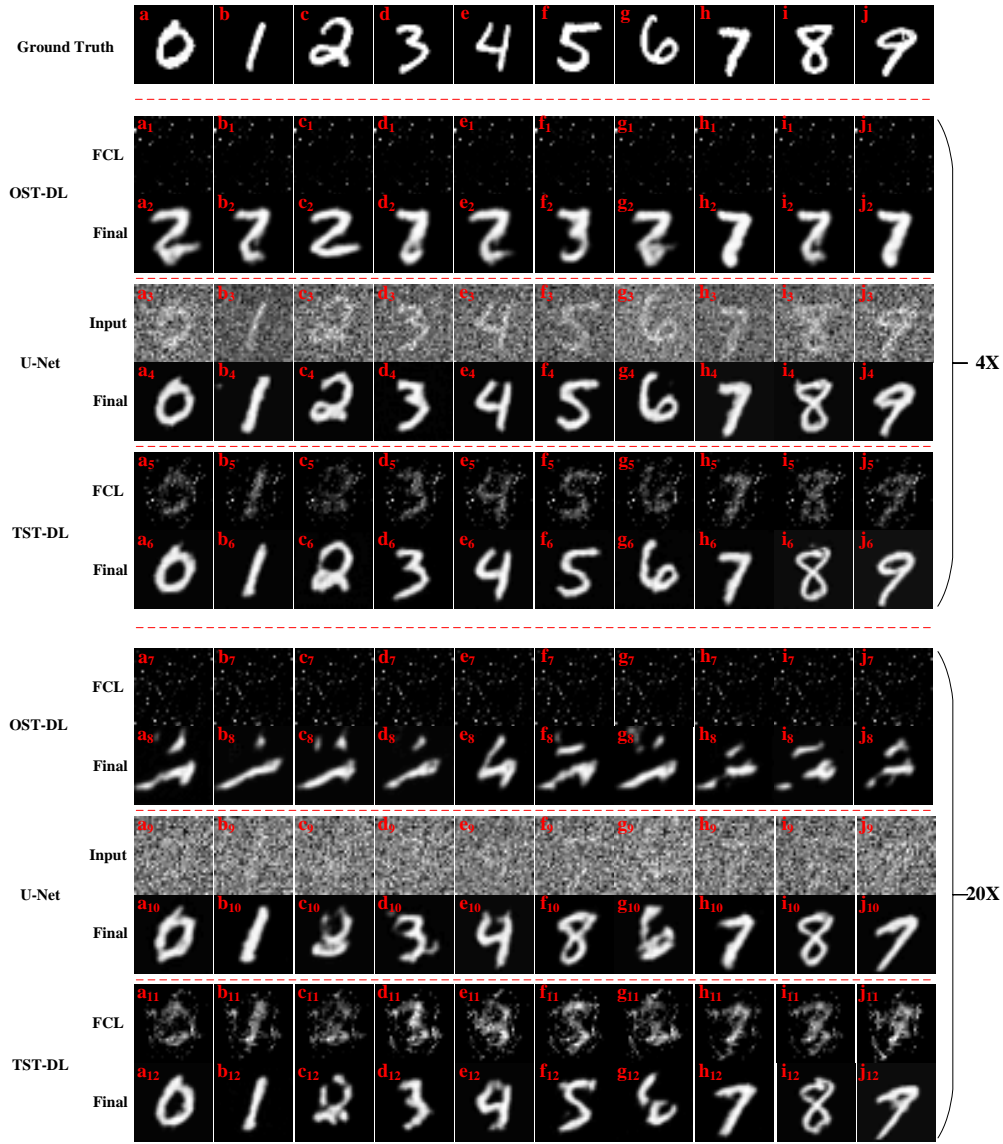

Fig. 11. Experimental results on single-pixel imaging with 4X and 20X compression ratios. (a-j) Ground-truth images. ($a_1$-$j_1$) Intermediate images after the FCL in OST-DL at the 4X compression ratio. ($a_2$-$j_2$) Final reconstructed images in OST-DL at the 4X compression ratio. ($a_3$-$j_3$) Initial image guesses as the inputs of U-Net at the 4X compression ratio. ($a_4$-$j_4$) Final reconstructed images in U-Net at the 4X compression ratio. ($a_5$-$j_5$) Intermediate images after the FCL in TST-DL at the 4X compression ratio. ($a_6$-$j_6$) Final reconstructed images in TST-DL at the 4X compression ratio. ($a_7$-$j_7$) Intermediate images after the FCL in OST-DL at the 20X compression ratio. ($a_8$-$j_8$) Final reconstructed images in OST-DL at the 20X compression ratio. ($a_9$-$j_9$) Initial image guesses as the inputs of U-Net at the 20X compression ratio. ($a_{10}$-$j_{10}$) Final reconstructed images in U-Net at the 20X compression ratio. ($a_{11}$-$j_{11}$) Intermediate images after the FCL in TST-DL at the 20X compression ratio. ($a_{12}$-$j_{12}$) Final reconstructed images in TST-DL at the 20X compression ratio.

## 5. Conclusions and discussion

A TST-DL framework is proposed for real-time computational imaging without prior knowledge of the imaging model while accounting for the input-output image size- and dimensionality- mismatch issues. The FCL in the first-step training directly learns the inverse of the forward operator given the training data. Then, the pre-trained FCL is fixed and concatenated with a U-Net for a second-step training as a regularization step. Simulations and experiments with different imaging models were conducted to verify the effectiveness of the proposed TST-DL with quantitative comparison with other DL frameworks and the iterative model-based optimization approaches. The average time to predict an image from the testing dataset in TST-DL is $\leq$ 1ms. The results show the TST-DL outperforms the other DL frameworks without physics priors and is comparable to (and sometimes better than) the DL framework and iterative optimization approaches that incorporate the known forward operator.

Although the TST-DL framework shows superior performance in these studies, there are still some trade-offs in the framework. A relatively large number of the training data is needed to optimize the parameters since the FCL is used to directly learn the model in the first step of TST-DL and it has a large number of parameters to train. Therefore, obtaining sufficient experimental training samples is a challenging issue although the problem occurs in most DL frameworks. In addition, training a large number of parameters also requires a significant amount of memory such that a powerful computer (such as a workstation) is a requirement while the commonly used DL frameworks with convolutional neural networks is applicable in a normal desktop with lower memory requirements. Besides, since a combination of DSSIM and RMSE is used as the loss function in the second-step training, the ratio between DSSIM and RMSE still needs to be determined heuristically based on the training dataset, although the performance was not very sensitive to changes in the ratio. Moreover, although it is shown that TST-DL can also reconstruct the images in nonlinear imaging models, further exploration is still needed in determining the optimal number of FCLs to use and the choose of the nonlinear activation functions in each FCL. Additionally, we would also like to introduce the uncertainly quantification in TST-DL to show the per-pixel accuracy of the reconstructed results in the future to better assess the confidence in the result.

In summary, the TST-DL approach enables reliable real-time image reconstruction without relying on possibly flawed assumptions about the imaging model. The two-step-training strategy constrains the training process so that the inverse model can be effectively learned. Overall, this provides a flexible, standardized framework that can be applied to diverse problems, including those with nonlinear models.

## 6. Funding, acknowledgments and disclosures

### 6.1 Funding

This work is supported by a Faculty Grant from the Neukom Institute for Computational Science at Dartmouth College.

### 6.2 Acknowledgments

We thank Dr. Shuming Jiao and Miss Jun Feng (Nanophotonics Research Center, Shenzhen University, China) for providing experimental data of single-pixel imaging.

### 6.2 Disclosures

The authors declare no conflicts of interest.